\documentclass[pra,twocolumn]{revtex4}

\usepackage{amsmath,amssymb,bbm,graphicx}

\begin{document}

\title{Quantizing Time}

\author{Kim Bostr\"om}
\affiliation{Institut f\"ur Physik, Universit\"at Potsdam, 
14469 Potsdam, Germany
}
\date{\today}

\begin{abstract}
    
A quantum mechanical theory is proposed which abandons an external parameter ``time'' in favor of a self-adjoint operator on a Hilbert space whose elements represent measurement events rather than system states. The standard quantum mechanical description is obtained in the idealized case of measurements of infinitely short duration. A theory of perturbation is developped. As a sample application Fermi's Golden Rule and the S-matrix are derived. The theory also offers a solution to the controversal issue of the time-energy uncertainty relation.

\end{abstract}

\maketitle

\section{Introduction}

Is there a need for quantizing time? If one is content with the current situation concerning the fundamental theories, then the answer is clearly ``no''. However, it turns out to be more and more difficult to finally merge the two most fundamental theories that we have, namely quantum theory and the theory of general relativity. If one takes the principle of relativity serious, then time and space should be treated on exactly the same footing. Quantum theory, however, treats time in a fundamentally different manner, namely as a classical parameter within a quantized theory. Even in quantum field theory the position of a particle is treated differently than the time of its detection: While the field operators $\hat\Psi(t,\boldsymbol x)$ are operator-valued distributions over $\boldsymbol x$, they are \emph{not} distributions over $t$. Rather, the time $t$ serves as a classical parameter for the continuous evolution of the system, just like in non-relativistic quantum mechanics. Therefore, in my point of view, as long as time is not treated \emph{exactly} on the same footing as position, the grand unification of fundamental theories will not succeed. 

In this paper a non-relativistic theory of one particle is presented where time and position both appear as ``canonical'' observables, i.e. as self-adjoint operators on a Hilbert space that are no functions of other operators.
This concept requires a modified axiomatic foundation which may seem unfamiliar to the reader. However, I have tried to keep the axiomatics as simple and intuitive as possible. The resulting theory includes a theory of measurement which yields the link between the abstract formalism and observations in the real world. The standard quantum theory turns out to be a special case where all measurements are performed within an infinitely short duration.
In order to show that the proposed theory is not only a kind of philosophical and formalistic gymnastics but also provides a toolbox for straight and transparent calculations, Fermi's Golden Rule and the S-Matrix in first order are calculated.

Now what is the problem with quantizing time? To put it shortly: A ``quantized'' quantity is represented by an observable; an observable represents a property of a physical system; the fundamental system of quantum mechanics is the particle; ``Time'' is not a property of particles. The problems and peculiarities arising with the construction of a time operator are the mathematical counterparts of this conceptual dilemma. 

Mathematically, by saying that ``position is a quantized quantity'' we mean that the wave functions $\psi(\boldsymbol x)=\langle\boldsymbol x|\psi\rangle$ are normalized over position space:
\begin{equation}
	\int d^3x\,|\psi(\boldsymbol x)|^2=1.
\end{equation}
The time-dependent wave functions $\psi(t,\boldsymbol x)$ are obtained with the help of the time evolution operator $\hat U(t,t_0)$:
\begin{equation}
	\psi(t,\boldsymbol x)=\langle\boldsymbol x|\hat U(t,t_0)|\psi\rangle.
\end{equation}
Because $\hat U(t,t_0)$ is unitary we have for all $t\in\mathbbm R$:
\begin{equation}\label{tint}
		\int d^3x\,|\psi(t,\boldsymbol x)|^2=1,
\end{equation}
Relation~(\ref{tint}) explicitely shows that time is not a quantized quantity, because the wave functions $\psi(t,\boldsymbol x)$ are normalized over space but not over time. 

Nonetheless, let us assume that there is an observable $\hat T$ measuring time in some way. Then we would expect that the corresponding Heisenberg operator $\hat T(t)$ obeys
\begin{equation}\label{Tdt}
	\frac d{dt}\hat T=\mathbbm1.
\end{equation}
Together with the Heisenberg equation of motion, relation~(\ref{Tdt}) implies that $\hat T$ is conjugate to the Hamiltonian $\hat H$ of the system,
\begin{equation}\label{TH}
    [\hat T,\hat H]=i\hbar.
\end{equation}
Defining an energy shift operator by
\begin{equation}
    \hat K(\eta):=e^{\frac{i}{\hbar}\hat T\eta},\quad \eta\in{\mathbbm R},
\end{equation}
then relation~(\ref{TH}) implies that $K(\eta)$ maps an eigenstate of energy $E$ onto an eigenstate of energy $E-\eta$:
\begin{equation}
    \hat K(\eta)|E\rangle=|E-\eta\rangle.
\end{equation}
Because $\eta$ is an arbitrary real parameter, the spectrum of $\hat H$ is ${\mathbbm R}$.
This contradicts the principle of the \emph{stability of matter} which demands that the energy spectrum must be bounded from below. 
Concluding, there is no self-adjoint operator $\hat T$ fulfilling~(\ref{TH}). This is the famous theorem by \emph{Wolfgang Pauli} \cite{Pauli26} who wrote:
\begin{quote}\em
We conclude that the introduction of an operator $T$ must fundamentally be abandoned and that the time $t$ in quantum mechanics has to be regarded as an ordinary real number.
\end{quote}
Anyway, during the almost 80 years between Pauli's theorem and today there have been numerous attempts to introduce an observable $\hat T$ with the dimension of time. These attempts can basically be classified into the following categories:
\begin{itemize}
\item
One constructs a self-adjoint operator $\hat T$ conjugate to a suitably defined \emph{unbound} pseudo-Hamiltonian $\hat{\cal H}$. There is no general rule for this procedure but there are certain examples that lead to physically sensible quantities, in particular so-called ``Arrival time'' operators. \cite{Delgado97,Muga98,Egusquiza99}.
\item
The condition for $\hat T$ to be self-adjoint is dropped. Instead, one looks for a maximally-symmetric operator $\hat T$ conjugate to $\hat H$ whose eigenvectors are non-orthogonal. The measurement of $\hat T$ is understood within the concept of generalized measurements via POVM (positive operator-valued measure) \cite{Busch01,Werner87}. 
\item
The original proof of Pauli has a loophole. Closer investigations show that in certain cases it is possible to find a self-adjoint operator $\hat T$ conjugate to $\hat H$ in spite of the semi-boundness of $\hat H$. However, the physical meaning of such an operator remains unclear \cite{Galapon99,Galapon01}.
\item
A clock system is considered which is described quantum mechanically and whose pointer value is related to the actual time \cite{Mandelstamm45,Hartle88}.
\end{itemize}
All these approaches are based on the assumption that the time operator is itself subject to time evolution. The quantity in whose respect this time evolution takes place is still a classical parameter. Therefore, one cannot speak of a ``quantization of time'' in the same sense as one speaks of a quantization of position. In contrast to that, the approach presented in this paper realizes a quantization of time in the very meaning of the word.

Closely related to the notion of a time operator is the \emph{time-energy uncertainty relation},
\begin{equation}\label{DEDt}
	\Delta E\Delta t\geq \hbar/2,
\end{equation}
which is also a controversal issue \cite{Aharonov60,Aharonov01,Busch01,Brunetti02,Eberly73}. 
There are theoretical considerations and experimental facts that strengthen an uncertainty relation between the measured energy of the system and the duration of this measurement. Such a time-energy uncertainty relation can also be derived in a quite general way when considering an interacting environmental system~\cite{Briggs00,Briggs01}.
On the other hand it has been shown~\cite{Aharonov60} that there are particular cases where the duration of an energy measurement does \emph{not} affect the uncertainty of the measurement result. This paradoxical situation can be solved~\cite{Aharonov01} if one distinguishes between the measurement of a  \emph{known} and an \emph{unknown} Hamiltonian: 
In case the Hamiltonian is unknown it must be \emph{estimated} by observing the dynamics of the system. The uncertainty of such estimation and the duration of the measurement obey a relation of the form~(\ref{DEDt}). Only if the Hamiltonian is explicitely known, then it is possible to design a measurement which circumvents a time-energy uncertainty relation. 
The peculiarities connected with the time-energy uncertainty relation and those connected with a time operator have a common cause, namely that time is still a \emph{classical} element within a quantized theory.

In this paper a framework is proposed in which time is described as a canonical quantum observable, i.e. by a self-adjoint operator on a Hilbert space which is not a function of other operators. Since this unavoidably conflicts with the basic concepts of quantum mechanics, we are forced to modify the axiomatic foundations of the standard theory so that time and space are treated \emph{on exactly the same footing}. The resulting framework will be referred to as ``Quantum Event Theory'', in short QET. Some consequences of the theory are:
\begin{itemize}
\item
The ordinary picture of a continuous time evolution is replaced by
a discrete sequence of single measurement events occurring in time and space.
\item
The theory is nonlocal in time, i.e. a particle is not in a certain ``state'' at a certain time.
\item
The dynamics is entirely driven by measurements. If there is no measurement then there is no dynamics.
\item
The dynamics is irreversible, hence one can define an ``arrow of time''.
\item
The energy of a particle is not represented by its Hamiltonian. Rather, the Hamiltonian can be used to \emph{estimate} the energy of the particle.
\item
Time and energy obey a canonical uncertainty relation, while the Hamiltonian does not obey an uncertainty relation with time. Thus if the Hamiltonian is explicitely known it can be measured without any restriction to the duration of the measurement to estimate the particle's energy.
\item
There is no ``measurement problem'' in that there is no unitary evolution interrupted by quantum jumps. 
\end{itemize}

\section{Quantum events}

A particle is actually not the thing we perceive in an experiment. What we perceive are \emph{measurement events} occurring in time and space which we \emph{relate} to the presence of a particle. Pointers, LED's, oscilloscopes, eyes, ears, these are all \emph{detectors} and they form the ``real world'' through our perception. The particle itself is just the imaginary object behind the detector events. Therefore, let us put forward the following postulate:
\begin{quote}\em
The fundamental object of Quantum Event Theory is not the particle but the event caused by a measurement on the particle.
\end{quote}
Let us introduce the ``\emph{elementary quantum event}''
\begin{equation}
	|t,\boldsymbol x\rangle,
\end{equation}
where $t\in\mathbbm R$, $\boldsymbol x\in\mathbbm R^3$. 
The object $|t,\boldsymbol x\rangle$, an abstract ``ket'', represents the occurence of a detector event (``click'') at time $t$ and at position $\boldsymbol x$. More precisely, $|t,\boldsymbol x\rangle$ represents the \emph{detection of a particle} at time $t$ and at position $\boldsymbol x$. Next consider the ``\emph{general quantum event}''
\begin{equation}\label{Psiket}
	|\Psi\rangle=\int d^4x\,\Psi(t,\boldsymbol x)|t,\boldsymbol x\rangle,
\end{equation}
where $\Psi(t,\boldsymbol x)$ is a complex-valued function over $\mathbbm R^4$ and $d^4x\equiv dt\,d^3x$ is the standard measure over the $\mathbbm R^4$. 
The ket $|\Psi\rangle$ represents an \emph{unsharp} detector event which takes place within the region in spacetime specified by the support of $\Psi(t,\boldsymbol x)$. Define the scalar product between two kets $|\Psi\rangle$ and $|\Phi\rangle$ through
\begin{equation}\label{scalarprod}
	\langle\Psi|\Phi\rangle:=\int d^4x\,
		\Psi^*(t,\boldsymbol x)\Phi(t,\boldsymbol x),
\end{equation}
and define the norm of the ket $|\Psi\rangle$ by
$\|\Psi\|:=\sqrt{\langle\Psi|\Psi\rangle}$,
then the space ${\cal E}_0$ of all kets with finite norm,
\begin{equation}
	{\cal E}_0:=\{|\Psi\rangle\mid\|\Psi\|<\infty\},
\end{equation}
is a Hilbert space which is isomorphic to the space $L^2(\mathbbm R^4)$ of square-integrable functions over spacetime,
\begin{equation}
	{\cal E}_0\simeq L^2(\mathbbm R^4).
\end{equation}
While the space ${\cal E}_0$ is an \emph{abstract} space, the space $L^2(\mathbbm R^4)$ is a \emph{function} space and contains ``\emph{event wave functions}''
\begin{equation}
	\Psi(t,\boldsymbol x):=\langle t,\boldsymbol x|\Psi\rangle
\end{equation}
which are spacetime representations of the abstract kets $|\Psi\rangle$. The isomorphism between both spaces is given by
\begin{equation}
	{\cal E}_0\ni|\Psi\rangle \leftrightarrow 
		\Psi(t,\boldsymbol x)\in L^2(\mathbbm R^4).
\end{equation}
Explicitely, any event wave function $\Psi(t,\boldsymbol x)$ is normalizable over spacetime:
\begin{equation}\label{txint}
	\int d^4x\,|\Psi(t,\boldsymbol x)|^2<\infty.
\end{equation}
Relation~(\ref{txint}) explicitely shows that time is now a quantized quantity. Comparing relations~(\ref{tint}) and~(\ref{txint}) we see that the event wave functions $\Psi(t,\boldsymbol x)$ cannot be interpreted as traditional wave functions. 
In order to give these objects a physical meaning we need a theory of \emph{measurement}, which will be done later on.

The ket representation~(\ref{Psiket}) is justified by the orthonormality and completeness relations
\begin{eqnarray}
	\langle t,\boldsymbol x|t',\boldsymbol x'\rangle
		&=&\delta(t-t')\delta(\boldsymbol x-\boldsymbol x')\\
	\int d^4x\,|t,\boldsymbol x\rangle\langle t,\boldsymbol x|
		&=&\mathbbm 1.
\end{eqnarray}
In this sense the ``\emph{spacetime basis}''
${\cal B} :=\{|t,\boldsymbol x\rangle\mid (t,\boldsymbol x)\in\mathbbm R^4\}$
is a complete and orthonormal basis for the Hilbert space ${\cal E}_0$.
Mathematically, the elementary quantum events $|t,\boldsymbol x\rangle$ are \emph{improper} (not normalizable) vectors outside the Hilbert space ${\cal E}_0$. They belong to the \emph{distribution space} $\Phi_0^\dag$ which is the dual to the \emph{test space} $\Phi_0\simeq{\cal S}(\mathbbm R^4)$, where ${\cal S}(\mathbbm R^4)$ is the \emph{Schwartz space} of rapidly decreasing functions over $\mathbbm R^4$.
While $\Phi_0,{\cal E}_0,\Phi_0^\dag$ are \emph{abstract} spaces, the spaces ${\cal S}(\mathbbm R^4),L^2(\mathbbm R^4)$ are \emph{function} spaces and the space ${\cal S}^\dag(\mathbbm R^4)$ is the space of linear-continuous \emph{functionals} over ${\cal S}(\mathbbm R^4)$.
Test space $\Phi_0$, Hilbert space ${\cal E}_0$ and distribution space $\Phi_0^\dag$ form a \emph{Gelfand triplet} $\Phi_0\subset{\cal E}_0\subset\Phi_0^\dag$. 

The event space ${\cal E}_0$ is the tensor product of the Hilbert space ${\cal H}_T\simeq L^2(\mathbbm R)$ spanned by the time eigenvectors and the standard Hilbert space ${\cal H}\simeq L^2(\mathbbm R^3)$ spanned by the position eigenvectors,
\begin{equation}
	{\cal E}_0={\cal H}_T\otimes{\cal H}.
\end{equation}
Let us introduce a convenient notation. Kets $|\cdot\rangle$ with a sharp edge belong to the full space ${\cal E}_0$ and kets $|\cdot)$ with a soft edge belong to a factor space of ${\cal E}_0$. For example, if $|t)$ is a time eigenvector then
\begin{equation}
	|\psi_t)=(t|\Psi\rangle
\end{equation}
is a vector from the standard Hilbert space ${\cal H}$. As a function of the time $t$, $|\psi_t)$ is called the ``\emph{time representation}'' of $|\Psi\rangle$. It has not to be confused with the ``trajectory'' of the particle.

In order to include spin degrees of freedom, the event space must be extended to the space
\begin{equation}
	{\cal E}_s:={\cal E}_0\otimes{\cal H}_s,
\end{equation}
where ${\cal H}_s$ ist the $(2s+1)$-dimensional Hilbert space containing the spin states of a spin-$s$ particle. For the sake of simplicity we will restrict our attention to the spin-0 case and consider only events in ${\cal E}_0$.

The kets $|t,\boldsymbol x\rangle$ are eigenvectors of the time and position operators $\hat t$, $\hat{\boldsymbol x}$, respectively, with
\begin{eqnarray}
	\hat t&:=&\int d^4x\,t\,|t,\boldsymbol x\rangle\langle t,\boldsymbol x|\\
	\hat{\boldsymbol x}&:=&\int d^4x\,\boldsymbol x\,
		|t,\boldsymbol x\rangle\langle t,\boldsymbol x|,
\end{eqnarray}
which are both essentially self-adjoint operators on ${\cal E}_0$.

Let the ``\emph{spacetime representation}'' $\acute A\equiv\acute A(t,\boldsymbol x)$ of an operator $\hat A$ be defined by
\begin{equation}
	\acute A(t,\boldsymbol x)\,\langle t,\boldsymbol x|
		:=\langle t,\boldsymbol x|\hat A,
\end{equation}
i.e.
\begin{equation}
	\langle t,\boldsymbol x|\hat A|\Psi\rangle
		=\acute A\,\Psi\equiv\acute A(t,\boldsymbol x)\Psi(t,\boldsymbol x).
\end{equation}
For example, the spacetime representations of $\hat t$ and $\hat{\boldsymbol x}$ read
\begin{equation}
	\acute t= t,\quad
	\acute{\boldsymbol x}=\boldsymbol x.
\end{equation}
Time and position are ``primary observables'', i.e. we do not have to \emph{explain} them. Everybody immediately \emph{knows} what they are, even \emph{animals} have an instinctive notion of time and position. On the other hand, energy and momentum are ``secondary observables''. It took the genius of \emph{Isaac Newton} to formulate the classical definition of energy and momentum. In quantum mechanics, energy and momentum can be introduced by \emph{de Brogli}'s concept of matter waves. A particle with energy $E$ and momentum $\boldsymbol p$ is associated with a plane wave
\begin{equation}
	\varphi_{E,\boldsymbol p}(t,\boldsymbol x)
		=e^{-\frac i\hbar(Et-\boldsymbol p\boldsymbol x)},
\end{equation}
with respective frequency and wavelength
\begin{eqnarray}
	\nu&=&\frac{E}{2\pi\hbar},\quad
	\lambda=\frac{2\pi\hbar}{|\boldsymbol p|},
\end{eqnarray}
propagating into the direction of $\boldsymbol p$.
Let us implement de Brogli's concept by introducing the ket
\begin{equation}
	|E,\boldsymbol p\rangle 
		:= \frac1{\sqrt{2\pi\hbar}^4}\int d^4x\,
		e^{-\frac i\hbar(Et-\boldsymbol p\boldsymbol x)}
		|t,\boldsymbol x\rangle,
\end{equation}
whose spacetime repesentation,
\begin{equation}
	\langle t,\boldsymbol x|E,\boldsymbol p\rangle
		=\frac1{\sqrt{2\pi\hbar}^4}\,
		e^{-\frac i\hbar(Et-\boldsymbol p\boldsymbol x)},
\end{equation}
is a plane wave. The prefactor has been chosen for normalization purposes, so that the set $\tilde{\cal B}=\{|E,\boldsymbol p\rangle\mid (E,\boldsymbol p)\in\mathbbm R^4\}$ represents an orthonormal basis for the event Hilbert space ${\cal E}_0$.
The kets $|E,\boldsymbol p\rangle$ are eigenvectors of the observables $\hat E$ and $\hat{\boldsymbol p}$ representing the energy and the momentum of the particle.
In spacetime representation $\hat E$ and $\hat{\boldsymbol p}$ read
\begin{eqnarray}
	\acute E&=&i\hbar\frac\partial{\partial t},\quad
	\acute{\boldsymbol p}
		=\frac\hbar i\boldsymbol\nabla.
\end{eqnarray}
By construction, energy and time as well as momentum and position are \emph{conjugate} to each other,
\begin{eqnarray}
	[\hat E,\hat t]&=&i\hbar,\quad
	[\hat{\boldsymbol p},\hat{\boldsymbol x}]=-i\hbar,
\end{eqnarray}
and we have the \emph{Fourier relations}
\begin{eqnarray}
	\Psi(t,\boldsymbol x)&=&\frac1{\sqrt{2\pi\hbar}^4}\int d^4p\,
		\tilde\Psi(E,\boldsymbol p)\,
		e^{-\frac i\hbar(Et-\boldsymbol p\boldsymbol x)}\\
	\tilde\Psi(E,\boldsymbol p)&=&\frac1{\sqrt{2\pi\hbar}^4}\int d^4x\,
		\Psi(t,\boldsymbol x)\,e^{\frac i\hbar(Et-\boldsymbol p\boldsymbol x)}.
\end{eqnarray}
Note that the energy $\hat E$ is not associated to the Hamiltonian, which has not yet been introduced into the theory. The Hamiltonian is a ``non-canonical'' observable, i.e. a nontrivial function of the canonical observables $\hat{\boldsymbol x}$ and $\hat{\boldsymbol p}$ and of the time parameter $t$.
The connection between the canonical energy $\hat E$ and the non-canonical Hamiltonian $\hat H$ yields the physical basis for the dynamics of the particle, as we will later see.

\section{Dynamics and measurement}\label{Dynamics}

Let us turn to the dynamics of the particle. The word ``dynamics'' suggests that there is something ``moving''. But with time as a quantum number, how can there be movement? Indeed, we will see that within the framework of QET there is no movement in the actual meaning of the word. Movement rather turns out to be an illusion, similiar to the fact that the pictures of a ``movie'' do not really move. Instead, a movie consists of a discrete sequence of \emph{stills} which generates the illusion of movement if it is played fast enough. QET describes an analogous situation: Instead of having a continuous \emph{particle trajectory} there is a sequence of quantum events which generates the illusion of a moving particle. One could say that QET describes reality as a ``quantum comic strip'' with something happening in each panel but nothing happening in between. Just like every comic strip has a \emph{plot} we must seek for a \emph{causal connection} between the events. For convenience let us stick to the term ``dynamics'' and understand it as the causal connection between subsequent events.

First we axiomatically introduce a ``propagator'' $\hat G$ which governs the dynamics of the particle. The explicit form of $\hat G$ will have to be derived from a physical postulate later on. 
So let there be an operator $\hat G$ on ${\cal E}$ called the ``\emph{propagator}''.
Define the ``\emph{orbit}'' of a quantum event $|\Psi\rangle\in{\cal E}$, $\langle\Psi|\Psi\rangle=1$, as its image under the propagator,
\begin{equation}
	|\psi\rangle := \hat G|\Psi\rangle.
\end{equation}
The ``\emph{orbit wave function}'',
\begin{equation}
	\psi(t,\boldsymbol x):=\langle t,\boldsymbol x|\hat G|\Psi\rangle,
\end{equation}
must not be confused with the \emph{event wave function} $\Psi(t,\boldsymbol x)=\langle t,\boldsymbol x|\Psi\rangle$. While $\Psi(t,\boldsymbol x)$ is normalizable over spacetime, the orbit wave function $\psi(t,\boldsymbol x)$, as we will later see, is \emph{not} normalizable over spacetime, so the vector $|\psi\rangle$ is in fact not a vector from the Hilbert space ${\cal E}_0$ but rather from the distribution space $\Phi_0^\dag$.

Any essentially self-adjoint operator $\hat A$ on ${\cal E}$ is called an ``\emph{observable}''.
An observable $\hat A$ that can be observed within an arbitrarily short period of time is called a ``\emph{instantaneous observable}'' and is of the form
\begin{equation}
	\hat A=\int dt\,|t)(t|\otimes\hat A(t),
\end{equation}
where $\hat A(t)$ indicates the explicit time-dependence of $\hat A$ and has not to be confused with the time evolution of $\hat A$ in the Heisenberg picture.
Only instantaneous observables commute with the time operator and can be used to prepare a ``quantum state'', because the concept of a state requires that the system shows the observed property at a given \emph{instance} in time. Consequently, all observables of standard quantum mechanics are instantaneous observables. In contrast to that, the canonical energy $\hat E$ is not instantaneous.

Define the ``\emph{operator density}'' of an observable $\hat A$ for a given orbit $|\psi\rangle=\hat G|\Psi\rangle$ as
\begin{eqnarray}\label{Aexp1}
	\langle\hat A\rangle_\psi(t,\boldsymbol x)
		&=&{\rm Re}\{\langle\psi|t,\boldsymbol x\rangle\langle t,\boldsymbol x|
		\hat A|\psi\rangle\}\\
		&=&\frac12\{\psi^*\acute A\psi+\psi\acute A^*\psi^*\}.
\end{eqnarray}
For example, 
\begin{eqnarray}
	\langle\hat{\boldsymbol x}\rangle_\psi(t,\boldsymbol x)
		&=&\boldsymbol x\,|\psi(t,\boldsymbol x)|^2\\
	\langle\hat{\boldsymbol p}\rangle_\psi(t,\boldsymbol x)
		&=&\frac {\hbar}{2i}\{\psi^*\boldsymbol\nabla\psi
		-\psi\boldsymbol\nabla\psi^*\}.\label{pdensity}
\end{eqnarray}
The density of the unity operator,
\begin{equation}
	\rho(t,\boldsymbol x):=\langle\mathbbm 1\rangle_\psi(t,\boldsymbol x)
		=|\psi(t,\boldsymbol x)|^2,\label{Nx}
\end{equation}
is called the ``\emph{particle density}''.

Now we come to the concept of \emph{measurements}. A measurement yields the link between abstract objects of the theory and phenomenons in the real world. Each measurement produces an event which is called the ``outcome'' of the measurement. In order to \emph{predict} the statistical distribution of measurement outcomes, we have to be given an \emph{initial event}. The event $|\Psi\rangle$ is the initial event of the orbit $|\psi\rangle=\hat G|\Psi\rangle$.

Measurements are performed within a certain spacetime region ${\cal W}\subset\mathbbm R^4$ which is called the ``\emph{observation window}''. A ``\emph{proper observation window}'' is defined as a region in spacetime where the number
\begin{eqnarray}
	W &:=&\int_{\cal W} d^4x\,\langle\mathbbm1\rangle_\psi(t,\boldsymbol x)
	\label{W}
\end{eqnarray}
is finite. Define the \emph{expectation value} of an observable $\hat A$ which is observed through the proper observation window ${\cal W}$ as
\begin{equation}
	\langle\hat A\rangle_\psi({\cal W}):=\frac1W\int_{\cal W}d^4x\,
		\langle\hat A\rangle_\psi(t,\boldsymbol x).\label{Aexp}
\end{equation}
The meaning of some important observables are:
\begin{itemize}
\item
$\mathbbm 1$ = ``Is the particle detected within the window?'' The answer is always ``yes'' because the window ${\cal W}$ serves as the \emph{statistical reference}. (The probability is $\langle\mathbbm 1\rangle_\psi({\cal W})=1$.) Events outside of ${\cal W}$ are not recognized by the statistics. In standard quantum mechanics the observation window is extended over the entire position space $\mathbbm R^3$ and is sharply peaked at a given time $t$.
\item
$\hat t$ = ``When is the particle detected within the window?'' The expression $\langle\hat t\rangle_\psi({\cal W})$ yields the expected time of detection within ${\cal W}$.
\item
$\hat{\boldsymbol x}$ = ``Where is the particle detected within the window?'' The expression $\langle\hat{\boldsymbol x}\rangle_\psi({\cal W})$ yields the expected position of a particle detected within ${\cal W}$.
\item
$\hat E$ = ``What is the energy of the particle when it is detected within the window?''. The expression $\langle\hat E\rangle_\psi({\cal W})$ yields the expected energy of a particle detected within ${\cal W}$.
\end{itemize}
A ``\emph{complete measurement}'' ${\cal M}$ is defined by an observation window ${\cal W}$ and a discrete set ${\cal P}=\{\hat\Pi_n\}$ of projectors on ${\cal E}$ such that
\begin{equation}
	\sum_n\hat\Pi_n=\mathbbm1,\quad\hat\Pi_n\hat\Pi_{m}=\delta_{nm}.
\end{equation}
Let the discrete set ${\cal A}=\{a_n\}$ contain the \emph{measurement results} of ${\cal M}$, then ${\cal M}$ corresponds to a measurement of the observable
\begin{equation}
	\hat A=\sum_n a_n\,\hat\Pi_n.
\end{equation}
The result $a\in{\cal A}$ occurs with the probability
\begin{equation}\label{Pa}
	P(a) =\langle\hat\Pi_a\rangle_\psi({\cal W}),
\end{equation}
where $\hat\Pi_a$ is the projector corresponding to the result ``$a$''.
By construction the probabilities sum up to unity,
\begin{eqnarray}
	\sum_a P(a)&=&\sum_a\langle\hat\Pi_a\rangle_\psi({\cal W})\\
		&=&\langle\sum_a\hat\Pi_a\rangle_\psi({\cal W})\\
		&=&\langle\mathbbm1\rangle_\psi({\cal W})=1,
\end{eqnarray}
thus $P(a)$ is a probability distribution on ${\cal A}$.
The occurence of the result ``$a$'' defines an event which is represented by the event vector
\begin{equation}\label{outcome}
	|\Psi_a\rangle:=\frac1{\sqrt{W P(a)}}\int_{\cal W}d^4x\,\hat\Pi_a
	|t,\boldsymbol x\rangle\langle t,\boldsymbol x|\psi\rangle,
\end{equation}
so that $\langle\Psi_a|\Psi_a\rangle=1$, as can be easily verified. The vector $|\Psi_a\rangle$ is the ``\emph{outcome}'' of ${\cal M}$.
\begin{figure}[h!]
	\[{\includegraphics[width=0.4\textwidth]{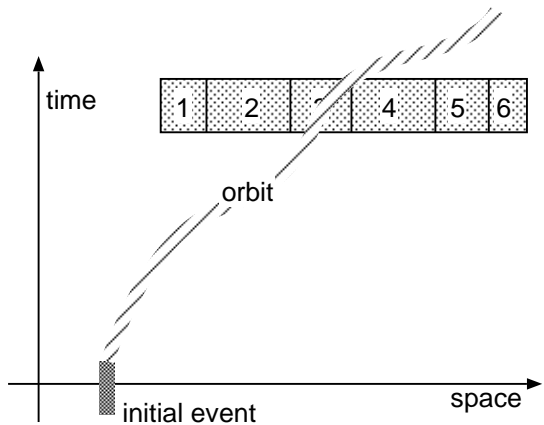}}\]
	\vspace*{-0.5cm}\caption{\small An initial event determines the probability distribution on the results of a complete observation. Here the position of the particle within a finite volume is measured (partitioned grey box). }\label{window}
\end{figure}

\section{Quantum history}

Consider a complete measurement ${\cal M}$ within the observation window ${\cal W}=T\times V$, where $T\subset\mathbbm R$ is the time interval and $V\subset\mathbbm R^3$ is the volume where the measurement takes place.
The measurement cannot be ``interrupted'' during the interval $T$; any interruption defines the end of the measurement and therefore $T$ reaches exactly up to this point. 
It is clear that \emph{during} the measurement ${\cal M}$ there cannot be another complete measurement on the same particle, otherwise both measurements would interfere with each other.
Thus all complete measurements on the same particle must occur in disjoint time intervals, and the outcome of each measurement determines the probability distribution on the outcomes of the next measurement. This \emph{next} measurement is defined as the ``later'' one.
If a sequence of complete measurements is carried out then the corresponding sequence of factual measurement results
\begin{equation}
	{\cal S} =(|\Psi_0\rangle,|\Psi_1\rangle,|\Psi_2\rangle,\ldots)
\end{equation}
represents the ``\emph{quantum history}'' of the particle, and
the time-ordering in ${\cal S}$ defines the arrow of time.
If the time intervals are short enough and follow each other closely enough, then this generates the illusion of a continuous evolution of the system. 

Eventually, it should be remarked that the so-called ``measurement problem'' does not appear within the framework of QET. There is no unitary time evolution interrupted by quantum jumps, because there is no continuous evolution: The orbit wave function  
$\psi(t,\boldsymbol x)=\langle t,\boldsymbol x|\hat G|\Psi\rangle$
is not a wave function in the traditional sense, i.e. a trajectory of amplitudes. 
The crucial point is that the initial event $|\Psi\rangle$ is in general \emph{smeared out in time}. 
There is no ``initial condition'', because there is no intial state, i.e. a set of properties which are measured exactly at a given time $t_0$. As a consequence there is no trajectory of states, i.e. we can no longer speak of a particle which is in a certain state at a certain time.

\begin{figure}[h!]
	\[{\includegraphics[width=0.4\textwidth]{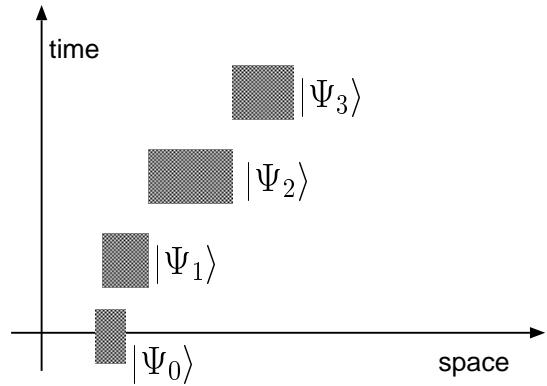}}\]
	\vspace*{-0.5cm}\caption{\small Complete measurements on one and the same particle have no time overlap. The result of one measurement yields the initial event for the next measurement. A discrete sequence of measurement events represents the ``quantum history'' of the particle. If time distances and uncertainties are small enough then the dynamics appears to be continuous.}\label{history}
\end{figure}

\section{The propagator}\label{Propagator}

So far we have axiomatically introduced the propagator $\hat G$ but we do not know how it acutally \emph{looks} like. Let us now derive the explicit form of the propagator from a physical postulate.

In classical mechanics the principle of the weakest action leads to Lagrangian mechanics and then via Legendre transformation to Hamiltonian mechanics. After applying the quantization rules (with all their ambiguities and difficulties) one ends up with quantum mechanics. However, time is still a classical parameter, so let us perform the final step and quantize time to end up with quantum event theory. The quantization rule for time is 
\begin{equation}\label{tquant}
	t\mapsto\hat t.
\end{equation}
Let us assume that for a one-particle system under study there is a Hamiltonian $\hat H=H(t,\hat{\boldsymbol x},\hat{\boldsymbol p})$ acting on the standard Hilbert space ${\cal H}$. Applying the quantization rule~(\ref{tquant}) the Hamiltonian becomes
\begin{equation}\label{H}
	\hat H=H(\hat t,\hat{\boldsymbol x},\hat{\boldsymbol p})
		=\int dt\,|t)(t|\otimes H(t,\hat{\boldsymbol x},\hat{\boldsymbol p}),
\end{equation}
which is an instantaneous observable on the event Hilbert space ${\cal E}_0$.
In order to relate $\hat H$ with the energy of the particle we postulate:
\begin{quote}\em
	The expectation value of the Hamiltonian coincides with the energy expectation value.
\end{quote}
Translated into formal language: For every observation window ${\cal W}\subset\mathbbm R^4$ the physically allowed orbit $|\psi\rangle=\hat G|\Psi\rangle$ of a particle should fulfill
\begin{equation}
	\langle\hat E\rangle_\psi({\cal W})=\langle\hat H\rangle_\psi({\cal W}).
\end{equation}
This implies that for all $(t,\boldsymbol x)\in\mathbbm R^4$
\begin{eqnarray}
	\langle\hat E\rangle_\psi(t,\boldsymbol x)
		&=&\langle\hat H\rangle_\psi(t,\boldsymbol x)\\
	\Leftrightarrow\quad
	{\rm Re}\{\langle\psi|t,\boldsymbol x\rangle\langle t,\boldsymbol x|
		\hat E|\psi\rangle\}
		&=&{\rm Re}\{\langle\psi|t,\boldsymbol x\rangle\langle t,\boldsymbol x|
		\hat H|\psi\rangle\},\nonumber \\
\end{eqnarray}
which implies that the orbit must fulfill
\begin{equation}
	\hat E|\psi\rangle=\hat H|\psi\rangle.\label{EH}
\end{equation}
Multiplying equation~(\ref{EH}) from the left with $(t|$, thus changing to time representation, we obtain the \emph{Schr\"odinger equation}
\begin{eqnarray}
	(t|\hat E|\psi\rangle&=&(t|\hat H|\psi\rangle\\
	\Leftrightarrow\quad
	i\hbar\frac\partial{\partial t}|\psi(t))&=&\hat H(t)|\psi(t)),
	\label{Schroedinger}
\end{eqnarray}
where $|\psi(t))\equiv(t|\psi\rangle$ is the time representation of the orbit. However, only in case of an initial event sharply peaked in time $|\psi(t))$ can be interpreted as the particle trajectory.
In this case we have an initial condition $|\psi(t_0))=|\psi_0)$ and everything looks familiar. However, in a realistic scenario there is no such initial condition because the initial event is not sharply peaked in time. Nonetheless, equation~(\ref{Schroedinger}) is still valid, but without $|\psi(t))$ referring to a trajectory of normalized states. 

A differential equation without initial condition is rather useless. All we can do then is calculate propagator matrix elements needed for expectation values.
Therefore, let us formulate a crucial condition for the propagator. Equation~(\ref{EH}) can be rewritten as
\begin{equation}
	\hat E\hat G|\Psi\rangle=\hat H\hat G|\Psi\rangle.
\end{equation}
The above equation is fulfilled for any initial event $|\Psi\rangle\in{\cal E}_0$ if the propagator obeys the condition
\begin{equation}\label{EHG}
	(\hat E-\hat H)\hat G=0.
\end{equation}
In time representation we get
\begin{eqnarray}
	(t|(\hat E-\hat H)\hat G|t')&=&0\\
	\left(i\hbar\frac\partial{\partial t}-\hat H(t)\right)
	\hat U(t,t')&=&0,\label{schroedU}
\end{eqnarray}
where we have set
\begin{equation}
	\hat U(t,t')\equiv (t|\hat G|t').
\end{equation}
Equation~(\ref{schroedU}) is nothing but the Schr\"odinger equation for the unitary time evolution operator $\hat U(t,t')$. Hence, the propagator is given by
\begin{equation}
	\hat G=\int dt\int dt'\,|t)(t'|\otimes \hat U(t,t').
\end{equation}
Another representation of $\hat G$ is
\begin{equation}
	\hat G=2\pi\hbar\,\delta(\hat E-\hat H).
\end{equation}
Using the relation
\begin{equation}
	\delta(x)=\frac i{2\pi}\left(\frac1{x+i\epsilon}
		-\frac 1{x-i\epsilon}\right),
\end{equation}
we can rewrite the propagator as
\begin{equation}
	\hat G=\hat G^++\hat G^-,
\end{equation}
where the Green operators
\begin{equation}
	\hat G^\pm:=\frac{\pm i\hbar}{\hat E-\hat H\pm i\epsilon}
\end{equation}
are the \emph{retarded and andvanced propagator}, respectively. This naming is justified as follows. Assume a time-independent Hamiltonian $\hat H$, then the retarded propagator can be written as
\begin{equation}
	\hat G^+=\int dE\,\frac{i\hbar}{E-\hat H+i\epsilon}|E)(E|.
\end{equation}
The time representation of $\hat G^+$ becomes
\begin{eqnarray}
	(t|\hat G|t_0)
		&=&\int dE\,\frac{i\hbar}{E-\hat H+i\epsilon}(t|E)(E|t_0)\\
		&=&\frac{i\hbar}{2\pi}\int dE\,\frac{1}{E-\hat H+i\epsilon}
		e^{-\frac i\hbar E(t-t_0)}\\
		&=&\theta(t-t_0)e^{-\frac i\hbar\hat H(t-t_0)},
\end{eqnarray}
where the residue theorem and the spectral decomposition of $\hat H$ has been used. Thus in fact the time representation of $\hat G^+$ coincides with the retarded time evolution operator
\begin{equation}
	(t|\hat G^+|t_0)=\hat U^+(t,t_0)=\theta(t-t_0)e^{-\frac i\hbar\hat H(t-t_0)}.
\end{equation}
In an analog way one verifies that
\begin{equation}
	(t|\hat G^-|t_0)=\hat U^-(t,t_0)=\theta(t_0-t)e^{-\frac i\hbar\hat H(t-t_0)}.
\end{equation}
So the retarded and advanced propagator $\hat G^\pm$ govern the dynamics to the future and the past, respectively.

\section{Finite observation time}

Does QET yield reasonable predictions for measurements of finite durarion? Such a scenario goes beyond the framework of standard quantum mechanics.

The initial event has the general form
\begin{equation}
	|\Psi\rangle=\int_T d\tau\,|\tau,\psi_\tau\rangle,
\end{equation}
where $T\subset\mathbbm R$ represents the time interval where the initial measurement takes place which produces $|\Psi\rangle$. The period $T$ may consist of several disjoint intervals, but the end points of $T$ always denote the beginning and the end of the entire measurement.
The vector $|\psi_\tau)\equiv(\tau|\Psi\rangle$ is \emph{not} a state trajectory, instead we have
\begin{eqnarray}
	\langle\Psi|\Psi\rangle=\int_T d\tau\,(\psi_\tau|\psi_\tau)=1.
\end{eqnarray}
The \emph{marginal operator density} of $\hat A$ at time $t$ is calculated by
\begin{eqnarray}
	\langle\hat A\rangle(t)&:=&\int d^3x\,
		\langle\hat A\rangle(t,\boldsymbol x).\label{Amarg}
\end{eqnarray}
The marginal particle density yields
\begin{eqnarray}
	\rho(t)&=&\langle\Psi|\hat G^\dag|t)(t|\hat G|\Psi\rangle\\
		&=&\int_T d\tau\int_T d\tau'\,(\psi_\tau|\hat U^\dag(t,\tau)
		\hat U(t,\tau')|\psi_{\tau'}).
\end{eqnarray}
Now we can use the group properties of the time evolution operator
\begin{eqnarray}
	\hat U^\dag(t,t')&=&\hat U(t',t),\\
	\hat U(t,t')\hat U(t',t'')&=&\hat U(t,t''),
\end{eqnarray}
to see that
\begin{eqnarray}
	\rho(t)&=&\int_T d\tau\int_T d\tau'\,
		(\psi_\tau|\hat U(\tau,\tau')|\psi_{\tau'})=N,\label{N}
\end{eqnarray}
where $N>0$ is a constant. For a window sharply peaked in time, ${\cal W}_t=[t-\epsilon,t+\epsilon]\times\mathbbm R^3$, we have
\begin{eqnarray}
	W&=&\int_{t-\epsilon}^{t+\epsilon}dt'\,\rho(t')
		=2\epsilon N.
\end{eqnarray}
The expectation value of of an observable $\hat A$ ``at time $t$'', i.e. within an infinitesimal interval around $t$, reads
\begin{eqnarray}
	\langle\hat A\rangle_t&:=&\langle\hat A\rangle({\cal W}_t)\\
		&=&\frac1{2\epsilon N}\int_{t-\epsilon}^{t+\epsilon}dt'\,
		\langle\hat A\rangle(t')\\
		&=&\frac1N\langle\hat A\rangle(t),\label{Aexpt}
\end{eqnarray} 
where $\langle\hat A\rangle(t)$ given by~(\ref{Amarg}) is the marginal operator density of $\hat A$ at time $t$.
Now let the window have finite duration, ${\cal W}=[t_1,t_2]\times\mathbbm R^3$, then
\begin{equation}
	W=\int_{t_1}^{t_2}dt'\,\rho(t')
		=(t_2-t_1)N.
\end{equation}
The expectation value of $\hat A$ through ${\cal W}$ reads
\begin{eqnarray}
    \langle\hat A\rangle_\psi({\cal W})
        &=&\frac1{(t_2-t_1)N}\int_{t_1}^{t_2}dt\,\langle\hat A\rangle(t)\\
		  &=&\frac1{t_2-t_1}\int_{t_1}^{t_2}dt\,\langle\hat A\rangle_t,
		  \label{Aexpfin}
\end{eqnarray}
which has the form of a \emph{time average} of the expectation value $\langle\hat A\rangle_t$ over the the time interval $[t_1,t_2]$. This is a reasonable result, although it has yet to be verified by experiments.

\section{Emergence of standard quantum mechanics}

In this section we consider a class of idealized measurements, namely measurements which are \emph{sharply peaked in time}, i.e. which have an infinitely short duration. We will see that for this special class of idealized measurements the formalism of QET becomes equivalent to the formalism of standard quantum mechanics. 

Let us assume that all measurements are performed within an infinitely short period of time. Let there be an initial measurement at time $t_0$ which serves as the preparation of an ``initial state'' $|\psi_0)\in{\cal H}$. The measurement result is represented by the initial event
\begin{equation}
	|\Psi\rangle=|t_0,\psi_0\rangle\equiv|t_0)\otimes|\psi_0),
\end{equation}
where $|t_0)$ is a time eigenvector and where we set $(\psi_0|\psi_0)=1$ for conveniency. Such an initial event is improper because it is not normalizable,
\begin{equation}
	\langle t_0,\psi_0|t_0,\psi_0\rangle=\delta(t_0-t_0)(\psi_0|\psi_0)
	=\delta(0).
\end{equation}
The time representation of the orbit $|\psi\rangle=\hat G|\Psi\rangle$,
\begin{eqnarray}
	|\psi(t))&:=&(t|\psi\rangle=(t|\hat G|\Psi\rangle
		=(t|\hat G|t_0,\psi_0\rangle\\
		&=&\hat U(t,t_0)|\psi_0)\label{traj}
\end{eqnarray}
is now indeed the \emph{trajectory} of a particle which evolves from the state $|\psi_0)\in{\cal H}$ at time $t_0$ to the state $|\psi(t))\in{\cal H}$ at time $t$. 
The orbit then has the form
\begin{eqnarray}
    |\psi\rangle&=&\hat G\,|\Psi\rangle=\hat G|t_0,\psi_0\rangle\\
        &=&\int dt\,|t)(t|\hat G|t_0,\psi_0\rangle\\
        &=&\int dt\,|t)\otimes \hat U(t,t_0)|\psi_0)\\
        &=&\int dt\,|t,\psi(t)\rangle,
\end{eqnarray}
and the spacetime representation of the orbit,
\begin{equation}
	\psi(t,\boldsymbol x)=\langle t,\boldsymbol x|\psi\rangle
		=(\boldsymbol x|\psi(t)),
\end{equation}
coincides with the familiar wave function of the particle.
The marginal density of an instantaneous observable 
\begin{equation}\label{Ainst}
	\hat A=\int dt\,|t)(t|\otimes\hat A(t)
\end{equation}
reads
\begin{eqnarray}
	\langle\hat A\rangle(t)
		&=&\int d^3x\,{\rm Re}\{\langle\psi|t,\boldsymbol x\rangle
		\langle t,\boldsymbol x|\hat A|\psi\rangle\}\\
		&=&{\rm Re}\{\langle\psi|t)(t|\hat A|\psi\rangle\}\\
		&=&(\psi(t)|\hat A(t)|\psi(t)).\label{At}
\end{eqnarray}
The marginal density of the particle number thus becomes
\begin{eqnarray}
	\rho(t)&=&\langle\mathbbm1\rangle(t)
		=(\psi(t)|\psi(t))\\
		&=&(\psi_0|\hat U^\dag(t,t_0)\hat U(t,t_0)|\psi_0).
\end{eqnarray}
Since $\hat U(t,t_0)$ is unitary we have
\begin{equation}
	\rho(t)\equiv 1.
\end{equation}
Consider an observation window which is sharply peaked at time $t$ and involves the entire position space $\mathbbm R^3$,
\begin{equation}\label{Wideal}
	{\cal W}_t:=[t-\epsilon,t+\epsilon]\times\mathbbm R^3,
\end{equation}
then by definition~(\ref{W})
\begin{eqnarray}
	W&=&\int_{t-\epsilon}^{t+\epsilon}dt'\int d^3x\,
		\langle\mathbbm 1\rangle(t',\boldsymbol x)
		=2\epsilon,
\end{eqnarray}
and thus the expectation value~(\ref{Aexp}) of the unstantaneous observable $\hat A$~(\ref{Ainst}) through this observation window is given by
\begin{eqnarray}
	\langle\hat A\rangle_t &:=&\langle\hat A\rangle({\cal W}_t)\\
		&=&\frac1{W}\int_{t-\epsilon}^{t+\epsilon}dt'
		\int d^3x\,\langle\hat A\rangle(t',\boldsymbol x)\\
		&=&\langle\hat A\rangle(t)
		=(\psi(t)|\hat A(t)|\psi(t)),\label{A_t}
\end{eqnarray}
where we have used~(\ref{At}). Expression~(\ref{A_t}) coincides with the expectation value of $\hat A$ at time $t$ in one-particle quantum mechanics.
The operator density of $\hat t$ yields
\begin{equation}
	\langle\hat t\rangle_\psi(t,\boldsymbol x)=t\,|\psi(t,\boldsymbol x)|^2,
\end{equation}
so the expectation value of $\hat t$, which measures the time when the particle is detected within the window ${\cal W}_t$, is given by
\begin{equation}
	\langle\hat t\rangle_t=\int d^3x\,t\,|\psi(t,\boldsymbol x)|^2=t.
\end{equation}
Thus the time operator $\hat t$ can be replaced by its expectation value $t$,
\begin{equation}
	\hat t\mapsto t.
\end{equation}
The probability of a measurement outcome ``$a_1$'' at time $t_1>t_0$ reads
\begin{equation}
	P(a_1,t_1)=\langle\hat\Pi_{a_1}\rangle_{t_1}
		=(\psi(t_1)|\hat\Pi_{a_1}|\psi(t_1)),
\end{equation}
where $\hat\Pi_{a_1}$ is the corresponding projector,
and the result of this outcome is represented by the vector
\begin{eqnarray}
	|\Psi_1\rangle&=&\frac1{\sqrt{ WP(a)}}\int_{t_1-\epsilon}^{t_1+\epsilon}dt
		\int d^3x\,\hat\Pi_{a_1}
		|t,\boldsymbol x\rangle\langle t,\boldsymbol x|\psi\rangle\\
		&=&\frac1{\sqrt{2\epsilon}}\int_{t_1-\epsilon}^{t_1+\epsilon}dt\,|t)
		\otimes\frac1{\sqrt{P(a_1)}}\hat\Pi_{a_1}(t)|\psi(t)).
\end{eqnarray}
For every finite $\epsilon$ this is a normalized vector concentrated around time $t_1$. In the limit $\epsilon\rightarrow0$ we replace it by the improper vector
\begin{equation}
	|\Psi_1\rangle=|t_1,\psi_1),
\end{equation}
where 
\begin{equation}
	|\psi_1)=\frac1{\sqrt{P(a_1)}}\hat\Pi_{a_1}(t_1)|\psi(t_1))
\end{equation}
is the ``post-measurement state'' at time $t_1$. This coincides with the projection postulate of standard quantum mechanics.
After sequential measurements at times $t_0,t_1, t_2,\ldots$, the quantum history of the particle is of the form
\begin{equation}
	{\cal S}=\{|t_0,\psi_0\rangle,|t_1,\psi_1\rangle,\ldots\},
\end{equation}
where the vectors $|\psi_0),|\psi_1),\ldots$ are normalized and can be interpreted as quantum states. 
If we like to, we may create the ``measurement problem'' by assuming that the particle evolves deterministically during the time $t\in(t_i,t_{i+1})$ along the ``trajectory''
\begin{equation}
	|\psi(t))=\hat U(t,t_i)|\psi_i),
\end{equation}
while at the time points $t=t_i$ it randomly ``jumps'' from one state to another with the probability $P(a_i)$:
\begin{equation}
	|\psi_i)\mapsto|\psi_{i+1}).
\end{equation}
Note that such construction is not possible in case of measurements of finite duration, because then there is no trajectory $|\psi(t))$.

Concluding, standard quantum mechanics is obtained from QET when instantaneous observables are measured through observation windows which are sharply peaked in time and include the entire position space $\mathbbm R^3$. 

\begin{figure}
	\[{\includegraphics[width=0.4\textwidth]{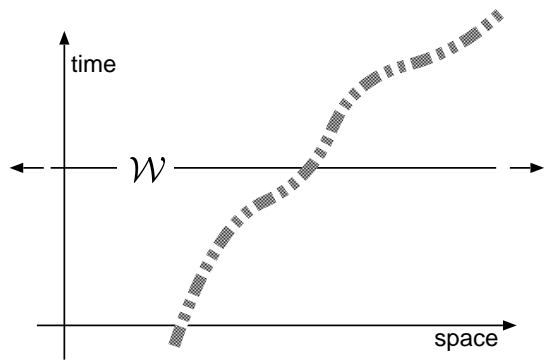}}\]
	\vspace*{-0.5cm}\caption{\small The observation window of standard quantum mechanics. The observation period is infinitesimally short and the observed volume is infinitely large.}\label{standard}
\end{figure}

\section{Continuity relation}

Let the Hamiltonian be of the form
\begin{equation}
	\hat H=\frac1{2m}\hat{\boldsymbol p}^2+\hat V,
\end{equation}
where the potential
\begin{equation}
	\hat V=\int d^4x\,V(t,\boldsymbol x)\,|t,\boldsymbol x\rangle
		\langle t,\boldsymbol x|
\end{equation}
is diagonal in time and space, hence it is an \emph{instantaneous} and \emph{local} potential.
The spacetime representation of $\hat H$ then reads
\begin{equation}
	\acute H=-\frac{\hbar^2}{2m}\Delta+V(t,\boldsymbol x).
\end{equation}
We have seen in section~\ref{Propagator} that the orbit fulfills the Schr\"odinger equation~(\ref{Schroedinger}), therefore the equation
\begin{equation}
	i\hbar\frac\partial{\partial t}\psi
		=\left(-\frac{\hbar^2}{2m}\Delta+V\right)\psi\label{schroed}
\end{equation}
holds even if $|\Psi\rangle$ is not sharply peaked in time, in which case there is no initial condition for $\psi$. Equation~(\ref{schroed}) implies the \emph{continuity relation}
\begin{equation}
	\frac\partial{\partial t}\rho+\boldsymbol\nabla\cdot\boldsymbol j=0,
\end{equation}
where
\begin{eqnarray}
	\rho(t,\boldsymbol x)&=&|\psi(t,\boldsymbol x)|^2,\\
	\boldsymbol j(t,\boldsymbol x)
		&=&\frac{\hbar}{2mi}\{\psi^*\boldsymbol\nabla\psi
		-\psi\boldsymbol\nabla\psi^*\}.
\end{eqnarray}
Using~(\ref{pdensity}) and~(\ref{Nx}) we can rewrite these expressions as
\begin{eqnarray}
	\rho(t,\boldsymbol x)&=&\langle\mathbbm 1\rangle(t,\boldsymbol x)\\
	\boldsymbol j(t,\boldsymbol x)
		&=&\langle\frac{\hat{\boldsymbol p}}{m}\rangle(t,\boldsymbol x),
\end{eqnarray}
If we reasonably consider the operator
\begin{equation}
	\hat{\boldsymbol v}:=\frac{\hat{\boldsymbol p}}{m}
\end{equation}
as the observable for the particle's velocity, then
the function $\boldsymbol j$ turns out as the \emph{velocity density} of the particle.
The continuity relation implies the existence of a ``conserved charge'' which is here nothing but the constant $N>0$ calculated in~(\ref{N}),
\begin{eqnarray}
	Q&:=&\int d^3x\,\rho(t,\boldsymbol x)
		=N.
\end{eqnarray}
If the initial event $|\Psi\rangle$ is sharply peaked in time then \mbox{$N=1$}. Only in this case $\rho(t,\boldsymbol x)$ can be interpreted as a \emph{probability density} and $\boldsymbol j(t,\boldsymbol x)$ as a \emph{probability current}. Note that the formalism of QET is not dependent of such a probability interpretation of $\rho$ and $\boldsymbol j$.

\section{Time-energy uncertainty}

An observable which is \emph{not} instantaneous is the energy $\hat E$. Let us explicitely calculate the marginal density of $\hat E$ at time $t$:
\begin{eqnarray}
	\langle\hat E\rangle(t)&=&\int d^3x\,\langle\hat E\rangle(t,\boldsymbol x)\\
		&=&\int d^3x\,{\rm Re}\{\langle\psi|t,\boldsymbol x\rangle
		\langle t,\boldsymbol x|\hat E|\psi\rangle\}\\
		&=&\int d^3x\,{\rm Re}\{(\psi(t)|\boldsymbol x)(\boldsymbol x|
		i\hbar\frac\partial{\partial t}|\psi(t))\}\\
		&=&(\psi(t)|\hat H(t)|\psi(t))
		=\langle\hat H\rangle(t),
\end{eqnarray}
where~(\ref{Schroedinger}) has been used.
Therefore the expectation values of energy $\hat E$ and Hamiltonian $\hat H$ in fact coincide at every instance in time,
\begin{equation}
	\langle\hat E\rangle_t=\langle\hat H\rangle_t,
\end{equation}
where~(\ref{Aexpt}) has been used.
But how about the \emph{variances} of $\hat E$ and $\hat H$?
For two self-adjoint operators $\hat A,\hat B$ the general uncertainty relation
\begin{equation}
	\Delta A\Delta B\geq\frac12|\langle[\hat A,\hat B]\rangle|
\end{equation}
holds, where the uncertainty of an observable $\hat A$ is defined as usual by
\begin{equation}
	\Delta A:=\sqrt{\langle\hat A^2\rangle-\langle\hat A\rangle^2}.
\end{equation}
Because time $\hat t$ and energy $\hat E$ are conjugate to another, $[\hat E,\hat t]=i\hbar$, there is a canonical time-energy uncertainty relation
\begin{equation}
	\Delta E\Delta t\geq\frac\hbar2,
\end{equation}
thus in fact $\hat E$ cannot be measured ``at time $t$'', i.e. within an infinitesimally short period, because the variance would be infinite. If instead $\hat E$ is measured within the finite period $[t_1,t_2]$, its expectation value reads
\begin{equation}
	\langle\hat E\rangle=\frac1{t_2-t_1}\int_{t_1}^{t_2} dt\,
		\langle\hat E\rangle_t,
\end{equation}
where we used~(\ref{Aexpfin}).
The measurement of $\hat E$ amounts to observing the system's dynamics which needs a finite observation period. 
On the other hand $\hat H$ is an instantaneous observable~(\ref{H}), so we have $[\hat H,\hat t]=0$, and thus
\begin{equation}
	\Delta H\Delta t=0.
\end{equation}
Therefore, the Hamiltonian of the particle can in principle be measured without any restriction to the duration of the measurement. 
Because $\hat H$ is itself not a canonical observable but rather a function of the canonical observables $\hat{\boldsymbol x}$ and $\hat{\boldsymbol p}$ and of the time parameter $t$, we need to \emph{know} its explicit form in order to measure it. 
It is this additional knowledge which is exploited when the energy $\hat E$ of the particle is estimated by a measurement of $\hat H$ without a restriction to the duration of the measurement.
If $\hat H$ is not known then one has to measure $\hat E$ in order to estimate $\hat H$. The shorter the measurement of $\hat E$, the more the result can deviate in each single measurement from the expectation value of $\hat H$. For very long measurements the measurement of $\hat E$ is a very good estimate for the Hamiltonian energy $\hat H$ and by that also for the energy $\hat E$ itself, because the expectation values of $\hat E$ and $\hat H$ coincide. This is the solution QET offers for the controversal issue of time-energy uncertainty. It goes well with the experimental facts and also with theoretical considerations based on standard quantum mechanics.
In particular cases the time-energy uncertainty principle has been verified experimentally. For instance, it is well known that if one attempts to measure the energetic state of an atom by observing absorbed or emitted radiatation, one is unavoidably confronted with a time-energy uncertainty relation. On the other hand, Aharonov et al. have showed~\cite{Aharonov60,Aharonov01} that there are experiments where the Hamiltonian can be measured in an arbitrarily short time, thus time-energy uncertainty appears to be violated. Such measurements, however, require that the Hamiltonian of the system is explicitely known. If the Hamiltonian is not known then it must be estimated by measurement. The uncertainty of the estimation and the duration of the measurement obeys a time-energy uncertainty relation.

\section{Perturbation theory}

In most applications the problem cannot be solved analytically. Therefore it is of significant use to have a theory of perturbation.
Since any calculation of expectation values involves the propagator $\hat G$, we need a perturbation series for this propagator in powers of a perturbation operator $\hat V$ which must be assumed as ``well behaved'' so that the series shows an acceptable convergence.

Let $\hat H_0$ be the Hamiltonian of the unperturbed system whose eigenvectors and eigenvalues are already known. 
Now add an instantaneous perturbation operator 
\begin{equation}
    \hat V=\int dt\,|t)(t|\otimes\hat V(t).
\end{equation}
so that the total Hamiltonian reads
\begin{eqnarray}
    \hat H&=&\hat H_0+\hat V.
\end{eqnarray}
Since the operators $\hat E $, $\hat H$ and $\hat H_0$ are self-adjoint, for any $\epsilon>0$ the operators 
\begin{eqnarray}
    \hat G^+&:=&\frac{i\hbar}{\hat E -\hat H+i\epsilon}\\
    \hat G_0^+&:=&\frac{i\hbar}{\hat E -\hat H_0+i\epsilon}
\end{eqnarray}
are well-defined.
For $\epsilon\rightarrow+0$ both operators become retarded Green operators, which only exist in a distributional sense.
Let us perform some algebraic transformations,
\begin{eqnarray}
    \hat E -\hat H_0-\hat V+i\epsilon&=&\hat E -\hat H+i\epsilon \\
    \hat E -\hat H_0+i\epsilon&=&\hat E -\hat H+i\epsilon+\hat V\\
    i\hbar&=&\hat G_0^+(\hat E -\hat H+i\epsilon)
        +\hat G_0^+\,\hat V
\end{eqnarray}
which, after application of $\hat G^+$ from the right and dividing by $i\hbar$, transforms into the recursive equation
\begin{equation}\label{Grecursive}
        \hat G^+=\hat G_0^+
            +\frac1{i\hbar}\hat G_0^+\,\hat V\,\hat G^+.
\end{equation}
This relation can be iterated giving the perturbation series
\begin{equation}\label{Gseries}
    \hat G^+=\hat G_0^+\sum_{n=0}^\infty
        \left\{\frac1{i\hbar}\hat V\hat G_0^+\right\}^n.
\end{equation}
For example, in first order the retarded propagator is approximated by
\begin{equation}\label{Gfirst}
    \hat G^+\approx \hat G_0^++\frac1{i\hbar}\hat G_0^+\hat V\hat G_0^+.
\end{equation}
Basically we are already done with perturbation theory. In order to see the advantage against standard quantum mechanics, let us look for the \emph{time representation} of the retarded propagator $\hat G^+$. We recall that
\begin{eqnarray}
    (t|\hat G^+|t_0)&=&\hat U^+(t,t_0)=\theta(t-t_0)\hat U(t,t_0)\\
    (t|\hat G_0^+|t_0)&=&\hat U_0^+(t,t_0)=\theta(t-t_0)\hat U_0(t,t_0).
\end{eqnarray}
Let us assume $t>t_0$ and insert $\int dt|t)(t|=\mathbbm 1$ into the recursive equation~(\ref{Grecursive}), which then becomes
\begin{widetext}
\begin{eqnarray}
    (t|\hat G^+|t_0)&=&(t|\hat G_0^+|t)
        +\frac1{i\hbar}(t_0|\hat G_0^+\,\hat V\,\hat G^+|t_0)\\
        &=&(t|\hat G_0^+|t)
        +\frac1{i\hbar}\int dt'\int dt''(t|\hat G_0^+|t')
        \hat V(t',t'')\,(t''|\hat G^+|t_0)
\end{eqnarray}
Because $\hat V(t',t'')\equiv(t'|\hat V|t'')=\delta(t'-t'')\hat V(t')$ we obtain
\begin{eqnarray}
    (t|\hat G^+|t_0)
        &=&(t|\hat G_0^+|t_0)
        +\frac1{i\hbar}\int dt'(t|\hat G_0^+|t')
        \hat V(t')\,(t'|\hat G^+|t_0)
\end{eqnarray}
and so, for all $t>t_0$,
\begin{eqnarray}
    \hat U^+(t,t_0)
        &=&\hat U_0^+(t,t_0)+\frac1{i\hbar}\int dt'\,
        \hat U_0^+(t,t')\hat V(t')\hat U^+(t',t_0)\\
    \hat U(t,t_0)
        &=&\hat U_0(t,t_0)+\frac1{i\hbar}\int_{t_0}^{t} dt'\,
        \hat U_0(t,t')\hat V(t')\hat U(t',t_0).\label{Upert}
\end{eqnarray}
Now we change to the \emph{interaction picture}. Here, the time evolution operator and the potential are respectively defined as
\begin{eqnarray}
    \hat U_I(t,t')&:=&\hat U_0^\dag(t,t')\hat U(t,t')
    =\hat U_0(t',t)\hat U(t,t')\\
    \hat V_I(t)&:=&\hat U_0^\dag(t,t_0)\,\hat V(t)\,\hat U_0(t,t_0).
\end{eqnarray}
Hitting~(\ref{Upert}) from the left with $\hat U_0^\dag(t,t_0)$ yields
\begin{eqnarray}
	\hat U_I(t,t_0)&=&\mathbbm 1+\frac1{i\hbar}\int_{t_0}^t dt'\,
		\hat U_0^\dag(t,t_0)\hat U_0(t,t')\hat V(t')\hat U(t',t_0)\\
		&=&\mathbbm1+\frac1{i\hbar}\int_{t_0}^t dt'\,
		\hat U_0(t_0,t')\hat V(t')\hat U(t',t_0)\\
		&=&\mathbbm1+\frac1{i\hbar}\int_{t_0}^t dt'\,
		\hat U_0^\dag(t',t_0)\hat V(t')
		\hat U_0(t',t_0)\hat U_0^\dag(t',t_0)\hat U(t',t_0)\\
		&=&\mathbbm1+\frac1{i\hbar}\int_{t_0}^t dt'\,
		\hat V_I(t')\hat U_I(t',t_0).
\end{eqnarray}
The above equation is just the integral form of the Schr\"odinger equation in the interaction picture,
\begin{equation}
	i\hbar\frac\partial{\partial t}\hat U_I(t,t_0)=\hat V_I(t)\hat U_I(t,t_0),
\end{equation}
together with the initial condition $\hat U_I(t_0,t_0)=\mathbbm1$. The solution is
\begin{eqnarray}
        \hat U_I(t,t_0)&=&\hat{\cal T}\left\{\sum_{n=0}^\infty\frac1{n!}
        \left(\frac1{i\hbar}\int_{t_0}^t dt'\,\hat V_I(t')\right)^n\right\}
        =\hat{\cal T}\left\{e^{-\frac i\hbar\int_{t_0}^t dt'\,
        \hat V_I(t')}\right\},
            \label{UDpert}
\end{eqnarray} 
where Dyson's \emph{time-ordering operator} $\hat{\cal T}$ cares for the correct ordering of the operators appearing in the above $n$-fold products,
\begin{equation}
    \hat{\cal T}\{\hat A(t_1)\hat B(t_2)\}
        :=\theta(t_1-t_2)\hat A(t_1)\hat B(t_2)
        +\theta(t_2-t_1)\hat B(t_2)\hat A(t_1),
\end{equation}
Concluding, the rather complicated expressions above are just different forms of the \emph{time representation} of the simple abstract relation~(\ref{Grecursive}). The time-ordering operator $\hat{\cal T}$ is not needed in QET, because the retarded propagator $\hat G^+$ \emph{tacitly} performs time-ordering.

\section{Fermi's Golden Rule}

As a sample application of QET's perturbation theory let us derive Fermi's Golden Rule. We will see that this is done in an intuitive and simple way. Let the free Hamiltonian have a mixed spectrum,
\begin{equation}
    \hat H_0=\sum_n \hbar\omega_n\,|n)(n|+\int_\sigma d\omega\,\rho(\omega)\,
    \hbar\omega\,|\omega)(\omega|,
\end{equation}
The energy density on the continuous spectrum $\sigma$ is given by $\rho(\omega)$ which vanishes outside of $\sigma$. The states are normalized according to
\begin{eqnarray}
    (n|m)=\delta_{nm},\quad (\omega|\omega')=\frac1{\rho(\omega)}\delta(\omega-\omega'),
	 \quad (n|\omega)=0.
\end{eqnarray}
The interaction potential couples the discrete levels to the continuum,
\begin{equation}
    \hat V=\sum_n\int d\omega\,\rho(\omega)\,V_n(\omega)\,|\omega)(n| + {\rm H.c.},
\end{equation}
where $V_n(\omega)=(\omega|\hat V|n)$.
In the remote past at $t=-T/2$ the system is in the discrete state $|n)$. Now we calculate the transition amplitude to a continuous state $|\omega)$ in the remote future at $t=T/2$ in first order perturbation theory,
\begin{eqnarray}
    \alpha_{n\rightarrow \omega}&\equiv&(\omega|\hat U(T/2,-T/2)|n)\\
        &=&\langle T/2,\omega|\hat G^+|\!-\!T/2,n\rangle\\
        &\approx& \langle T/2,\omega|\hat G_0|\!-\!T/2,n\rangle
        +\langle T/2,\omega|\hat G_0^+\hat V\hat G_0^+|\!-\!T/2,n\rangle,
\end{eqnarray}
where we made use of~(\ref{Gfirst}).
There is no zeroth-order transition from discrete to continuous levels,
\begin{eqnarray}
    \langle t,\omega|\hat G_0^+|t',n\rangle
        &=&\int dE\,\frac{i\hbar}{E-\hbar\omega_n+i\epsilon}
        \langle t,\omega|E)(E|t',n\rangle\\
        &=&\int dE\,\frac{i\hbar}{E-\hbar\omega_n+i\epsilon}
        \frac1{2\pi\hbar}e^{-\frac i\hbar E(t-t')}(\omega|n)
        =0.
\end{eqnarray}
Between the discrete levels we have a zeroth-order transition amplitude of
\begin{eqnarray}
    \langle t,m|\hat G_0^+|t',n\rangle
        &=&\int dE\,\frac{i\hbar}{E-\hbar\omega_n+i\epsilon}
        \langle t,m|E)(E|t',n\rangle\\
        &=&\int dE\,\frac{i\hbar}{E-\hbar\omega_n+i\epsilon}
        \frac1{2\pi\hbar}e^{-\frac i\hbar E(t-t')}(m|n)\\
        &=&\theta(t-t')\,\delta_{nm}\,e^{-i\omega_n(t-t')},
\end{eqnarray}
and for the continuous levels we find
\begin{eqnarray}
    \langle t,\omega|\hat G_0^+|t',\omega'\rangle
	 	&=&\int dE\,\frac{i\hbar}{E-\hbar\omega'+i\epsilon}
        \langle t,\omega|E)(E|t',\omega'\rangle\\
        &=&\int dE \,\frac{i\hbar}{E-\hbar\omega'+i\epsilon}
        \frac1{2\pi\hbar}e^{-\frac i\hbar E(t-t')}(\omega|\omega')\\
        &=&\theta(t-t')\,\frac1{\rho(\omega)}\delta(\omega-\omega')\,
        e^{-i\omega(t-t')},
\end{eqnarray}
respectively.
Thus we obtain
\begin{eqnarray}
    \alpha_{n\rightarrow \omega}&\approx&0+\frac1{i\hbar}
        \langle T/2,\omega|\hat G_0^+\hat V\hat G_0^+|\!-\!T/2,n\rangle\\
        &=&\frac1{i\hbar}\int dt\sum_m\int d\omega'\,\rho(\omega')\,
        \langle T/2,\omega|\hat G_0^+|t,\omega'\rangle\,V_m(\omega')\,
        \langle T/2,m|\hat G_0^+|t,n\rangle\\
        &=&\frac1{i\hbar}\int_{-T/2}^{T/2}dt\sum_m\int d\omega'\,V_m(\omega')
        \delta(\omega-\omega')e^{-i\omega(T/2-t)}\,
        \delta_{nm}e^{-i\omega_n(t+T/2)}\\       
        &=&\frac1{i\hbar}e^{-i(\omega_n+\omega)T/2}\,V_n(\omega)
        \int_{-T/2}^{T/2}dt\,e^{i(\omega-\omega_n)t}.
\end{eqnarray}
Now let us investigate the corresponding transition probability,
\begin{eqnarray}
    P_{n\rightarrow \omega}&=&|\alpha_{n\rightarrow \omega}|^2\\
        &\approx&\frac1{\hbar^2}\,|V_n(\omega)|^2
        \int_{-T/2}^{T/2}dt\,e^{i(\omega-\omega_n)t}
        \int_{-T/2}^{T/2}dt'\,e^{-i(\omega-\omega_n)t'}.
\end{eqnarray}
Since $T$ is very large, we can approximate the first integral by $2\pi\delta(\omega-\omega_n)$ and insert the peak into the second integral, which leads to
\begin{eqnarray}
    P_{n\rightarrow \omega}
	 	&\approx&\frac{2\pi T}{\hbar}|V_n(\omega)|^2\delta(\omega-\omega_n).
\end{eqnarray}
The \emph{transition rate}, defined by
\begin{equation}
    \Gamma_n(\omega):=\frac{P_{n\rightarrow \omega}}{T},
\end{equation}
becomes
\begin{equation}
    \Gamma_n(\omega)=\frac{2\pi}{\hbar}\delta(\omega-\omega_n)|V_n(\omega)|^2,
\end{equation}
The total transition rate out of the discrete level $|n)$ is then obtained by integration over the continuous spectrum,
\begin{eqnarray}
    \Gamma_n&=&\int_\sigma d\omega\,\rho(\omega)\,\Gamma_n(\omega)
        =\frac{2\pi}{\hbar}\rho(\omega_n)|V_n(\omega_n)|^2,
\end{eqnarray}
which coincides with Fermi's Golden rule.

\section{Scattering theory}

As another application let us consider a scattering scenario. Here the free Hamiltonian has the form
\begin{equation}
    \hat H_0=\frac1{2m}\hat{\boldsymbol p}^2,
\end{equation}
and the interaction potential is assumed to be constant in time,
\begin{equation}
    (t|\hat V|t')=\delta(t-t')\hat V.
\end{equation}
The scattering matrix elements are the transition amplitudes between plane waves $|\boldsymbol p')$ and $|\boldsymbol p)$ in the remote past and future, respectively, 
\begin{eqnarray}
    S(\boldsymbol p,\boldsymbol p')
        &:=&(\boldsymbol p|\hat U(T/2,-T/2)|\boldsymbol p')\\
        &=&\langle T/2,\boldsymbol p|
        \,\hat G^+\,|\!-\!T/2,\boldsymbol p'\rangle,
\end{eqnarray}
where $T$ is assumed to be very large.
Using the perturbative expansion~(\ref{Gseries}) one can derive the scattering matrix to any desired order. As an example let us calculate the S-matrix in first order from~(\ref{Gfirst}). 
Setting $\omega_{\boldsymbol p}:=\frac{\boldsymbol p^2}{2m}$ we have
\begin{eqnarray}
    \langle t,\boldsymbol p|\hat G_0^+|t',\boldsymbol p'\rangle
        &=&\int dE\,\frac{i\hbar}{E-\hbar\omega_{\boldsymbol p}+i\epsilon}
        \langle t,\boldsymbol p|E)(E|t',\boldsymbol p'\rangle\\
        &=&\int dE\,\frac{i\hbar}{E-\hbar\omega_{\boldsymbol p}+i\epsilon}
		  \frac1{2\pi\hbar}
        e^{-\frac i\hbar E(t-t')}(\boldsymbol p|\boldsymbol p')\\
        &=&\theta(t-t')\,\delta(\boldsymbol p-\boldsymbol p')\,
        e^{-i\omega_{\boldsymbol p}(t-t')}.
\end{eqnarray}
Using this together with the unity decomposition
\begin{equation}
    \mathbbm 1=\int dt\int d^3p\,|t,\boldsymbol p\rangle\langle t,\boldsymbol p|,
\end{equation}
we derive
\begin{eqnarray}
    \langle T/2,\boldsymbol p|\hat G_0^+\hat V\hat G_0^+
    |\!-\!T/2,\boldsymbol p'\rangle
    &=&\int dt\int d^3p''\int dt'\int d^3p'''\,
    \langle T/2,\boldsymbol p|\hat G_0^+|t,\boldsymbol p''\rangle
    \langle t,\boldsymbol p''|\hat V
	 |t',\boldsymbol p'''\rangle\times\nonumber \\
    &&\times\,
    \langle t',\boldsymbol p'''|\hat G_0^+|\!-\!T/2,\boldsymbol p'\rangle\\
    &=&\int dt\int d^3p''\int dt'\int d^3p'''\,
    \theta(T/2-t)\delta(\boldsymbol p-\boldsymbol p'')
    e^{-i\omega_{\boldsymbol p}(T/2-t)}\times\nonumber \\
    &&\times\,\delta(t-t')V(\boldsymbol p'',\boldsymbol p''')
    \theta(t'+T/2)\delta(\boldsymbol p'''-\boldsymbol p')
    e^{-i\omega_{\boldsymbol p'}(t'+T/2)}\\
    &=&e^{-\frac i\hbar(\omega_{\boldsymbol p}+\omega_{\boldsymbol p'})T/2}
    V(\boldsymbol p,\boldsymbol p')\int_{-T/2}^{T/2}dt\,
    e^{\frac i\hbar t(\omega_{\boldsymbol p}-\omega_{\boldsymbol p'})},
\end{eqnarray}
where $V(\boldsymbol p,\boldsymbol p')\equiv(\boldsymbol p|\hat V|\boldsymbol p')$. Since $T$ is very large, the integral can be approximated by a $\delta$-function peaked around $\omega_{\boldsymbol p}$, so that
\begin{eqnarray}
    \langle T/2,\boldsymbol p|\hat G_0^+\hat V\hat G_0^+
    |\!-\!T/2,\boldsymbol p'\rangle
    &\approx&e^{-i\omega_{\boldsymbol p}T}\,2\pi\hbar\,
    \delta(\omega_{\boldsymbol p}-\omega_{\boldsymbol p'})
    (\boldsymbol p|\hat V|\boldsymbol p').
\end{eqnarray}
Now that we have gathered all necessary information we are able to calculate the S-Matrix in first order as
\begin{eqnarray}
    S(\boldsymbol p,\boldsymbol p')
        &\approx&\langle T/2,\boldsymbol p|\big\{\hat G_0^+
        +\frac1{i\hbar}
        \hat G_0^+\hat V\hat G_0^+\big\}|\!-\!T/2,\boldsymbol p\rangle\\
        &\approx&e^{-i\omega_{\boldsymbol p}T}\big\{
        \delta(\boldsymbol p-\boldsymbol p')
        -2\pi i\,\delta(\omega_{\boldsymbol p}-\omega_{\boldsymbol p'})\,
        (\boldsymbol p|\hat V|\boldsymbol p')\big\}.
\end{eqnarray}
In standard quantum mechanics, the S-matrix is usually calculated in the interaction picture, where the irrelevant term $e^{-i\omega_{\boldsymbol p}T}$ vanishes. 

\end{widetext}

\section{Conclusion and Outlook}

A quantum mechanical theory has been proposed where time is treated as a self-adjoint operator rather than a real parameter. The fundamental objects of the theory are measurement events instead of particle states, which justifies the name ``Quantum Event Theory'', in short QET.
A theory of measurement has been given which allows an intuitive physical interpretation of the abstract formalism. The predictions of QET coincide with the predictions of standard quantum mechanics in the case of measurements of infinitely short duration. For measurements of finite duration the predictions of QET appear to be physically reasonable but have yet to be verified experimentally. A theory of perturbation has been developped which allows to calculate expectation values and transition amplitudes to any desired order. The calculations are transparent and are not more complicated to carry out as compared to standard quantum mechanics. In order to show this, the formalism of QET has been applied to derive Fermi's Golden Rule and the scattering matrix in first order.

Within the framework of QET the dynamics of a particle is driven by measurements only. Neither is there a ``state'' of the particle at a certain time, nor is there a ``trajectory'', i.e. a family of states parametrized by time. The evolution of the particle is represented by its ``quantum history'' ${\cal S}$, which is a discrete sequence of factual events that are probabilistic outcomes of complete measurements on the particle. Any physically allowed measurement sequence can be time-ordered, and the probability for each measurement result depends on the result of the preceding meaurement. This allows to define an arrow of time. 

The energy of the particle is represented by a canonical operator rather than by the Hamiltonian which can, however, be used to estimate the particle's energy. Time and energy obey a canonical commutation relation which directly leads to a time-energy uncertainty relation. There is no such relation between time and Hamiltonian, so that in case the Hamiltonian is explicitely known, it can be used to estimate the energy of the particle by a measurement of arbitrarily short duration. This offers a solution to the controversy about the time-energy uncertainty relation.

It would be interesting to apply QET to other quantum mechanical problems involving measurements of time, e.g. the tunneling time of a particle through a potential barrier.
Also, one could further explore the philosophical implications of QET concerning our understanding of time. 
Finally, it would be desirable to find a relativistic formulation of QET. 
Maybe some of the infinitites that quantum field theory is plagued with disappear when the field operators are generated by initial events which are smeared out in time.

This work is supported by the Deutsche Forschungsgemeinschaft (DFG).

\end{document}